\documentclass[showpacs,amsmath,nofootinbib,amssymb,epsfig]{revtex4}
\pdfoutput=1
\usepackage{graphicx} %Include figure files
\usepackage{dcolumn}% Align table columns on decimal point
\usepackage{bm}% bold math
\usepackage{caption}
\usepackage{subfig}

\usepackage{color}

\usepackage{amssymb}

\begin{document}

\title{Inflation via logarithmic entropy-corrected holographic dark energy model }
\author{F. Darabi }\email{f.darabi@azaruniv.ac.ir}
\author{F. Felegary}\email{falegari@azaruniv.ac.ir}
\affiliation{Department of Physics, Azarbaijan Shahid Madani University, Tabriz, 53714-161 Iran}
\author{M. R. Setare }\email{rezakord@ipm.ir}
\affiliation{Department of Science, Campus of Bijar, University of Kurdistan, Bijar , Iran}

\date{\today}

\begin{abstract}

 We study the inflation via logarithmic entropy-corrected holographic dark energy LECHDE model with future event horizon, particle horizon and Hubble horizon cut-offs, and compare the results with those of obtained in the study of inflation by holographic dark energy HDE model. In comparison,
the spectrum of primordial scalar power spectrum in the LECHDE model becomes redder than the spectrum in HDE model. Moreover, the consistency with the observational data in LECHDE model of inflation, constrains the reheating temperature and Hubble parameter by one parameter of holographic dark energy and two new parameters of logarithmic corrections.
\end{abstract}
\vspace{1cm}
\maketitle

\section{Introduction}
The recent cosmological and astrophysical data from Cosmic Microwave Background
radiation (CMB), the observations type Ia supernovae and Large Scale Structure
(LSS) persuasive express that the universe experiences an accelerated expansion phase \cite{Riess}. The accelerated expansion phase is derived by a energy component with negative pressure so called dark energy (DE). The most simple
candidate for dark energy is the cosmological constant, $\Lambda$. However,
the cosmological constant candidate suffers from the fine-tuning
and the cosmic coincidence problems  \cite{copeland,Swein}. Therefore, cosmologists suggested some  different models for DE  including tachyon, quintessence, phantom, k-essence, chaplygin gas, holographic and new agegraphic models \cite{Li1,Li2,wei,pad,setare,msetare,MRsetare}.

The holographic dark energy model HDE is one of the models of quantum gravity.
This model, based on the holographic principle, was proposed in Ref.\cite{suss}
by introducing the following energy density\begin{equation}
\varrho_{\Lambda}=3c^{2}M_{P}^{2}L^{-2},
\end{equation}
where $c$ is a numerical constant to be determined by observational data. $L$ and $M_{P}$ are the cut-off radius and the reduced Planck
mass, respectively.

The Bekenstein-Hawking entropy $S_{BH}=\frac{A}{4G}$ which is satisfied on the horizon, plays a fundamental
role in the HDE model \cite{wald}.
In fact,  $A\sim L^{2}$ is the area of
 horizon and since the holographic dark energy model is related to the area
law of entropy, therefore any correction to the area law of entropy will
modify the energy density of the HDE model.
One correction to the area law of entropy is the logarithmic correction \cite{Maj}
\begin{equation}
S_{BH}=\frac{A}{4G}+\tilde{\alpha}\ln(\frac{A}{4G})+\tilde{\beta}.
\end{equation}
Here $\tilde{\alpha}$ and $\tilde{\beta}$ are dimensionless constants.
The correction terms play a  fundamental role in the early time inflation
and late-time acceleration of the universe \cite{cai}. The corresponding
modified energy density of the Logarithmic Entropy-Corrected Holographic Dark Energy (LECHDE
) model has expressed by Wei \cite{WWi}
\begin{equation}
\rho_{\Lambda}=3c^{2}M_{P}^{2}L^{-2}+L^{-4}\Big[
\alpha \ln(M_{P}^{2}L^{2})+\beta\Big],\label{density}
\end{equation}
where $\alpha$ and $\beta$ are dimensionless constants.
In Eq. (\ref{density}), the second and third terms are comparable to the
first term when $L$ takes a very small amount. This means that the correction terms are important in early  universe and when the universe becomes large,
the second and third terms are ignorable and the logarithmic entropy-corrected holographic dark energy model reduces to the ordinary holographic dark energy model. The fractional energy density of LECHDE is given by
\begin{equation}
\Omega_{\Lambda}=\frac{\rho_{\Lambda}}{3M_{P}^{2}H^{2}}.\label{Omega}
\end{equation}

The holographic dark energy model was introduced to account for the present acceleration of the universe at low energy scale. However, by imposing the quantum gravity corrections to this model which led to LECHDE model
we are inevitably concerned with high energy state of the universe, namely
inflation. Inflation is the principal theoretical framework which describes the very early universe. In this work our aim is to study the effect of
logarithmic entropy-corrected holographic dark energy model on the inflation and the Cosmic Microwave Background spectrum (CMB).

We emphasize that
the study of holographic dark energy model, considering the cosmological constant problem,
leads to the fact that the Hubble horizon and particle horizon cut-offs contradict observations, and only the one with the future event horizon cut-off is consistent with  observations \cite{Li2}. However,
for the sake of generality, in this work we intend to study the inflation with logarithmic entropy-corrected holographic dark energy model considering the future event horizon, particle horizon and Hubble horizon cut-offs. 
%-----------------------------------------------------------------------------------------------
\section{  inflation and perturbational analysis }

In this section we study the inflation derived by a single minimally coupled
inflaton field. The energy density of inflaton field is given
by \cite{chen}
\begin{equation}
\rho_{\varphi}=\frac{1}{2}\dot{\varphi}^{2}+V(\varphi),\label{varphi}
\end{equation}
where $\varphi$ is the inflaton field and $V(\varphi)$ is the inflaton potential.
For simplicity, we assume that the inflaton field does not couple to the logarithmic
entropy-corrected holographic dark energy. Therefore, we can write the equation
of motion of the inflaton field, without affecting by the existence of the
logarithmic entropy-corrected holographic dark energy, as
\begin{equation}
\ddot{\varphi}+3H\dot{\varphi}+V_{\varphi}=0,\label{motion}
\end{equation}
where $V_{\varphi}=\frac{dV}{d\varphi}$. Moreover, we consider the slow-roll conditions \cite{chen}
\begin{equation}
\delta\equiv-\frac{\ddot{\varphi}}{H\dot{\varphi}},~~~~~~~~~~~~|\delta|\ll1,
\label{delta}
\end{equation}
\begin{equation}
\epsilon\equiv-\frac{\dot{H}}{H^{2}},~~~~~~~~~~~~|\epsilon|\ll1. \label{epp}
\end{equation}
We assume that the reheating period occurs immediately after the inflation
period. So, the number of e-folding is given by \cite{david}
\begin{equation}
N_{COBE}=62-\ln\Big(\frac{10^{16}GeV}{V_{end}^{\frac{1}{4}}}\Big)
-\frac{1}{3}\ln\Big(\frac{V_{end}^{\frac{1}{4}}}{T_{reh}^{\frac{1}{4}}}\Big),
\end{equation}
where $T_{reh}$ is the reheating temperature and $V_{end}$ is the potential
corresponding to the end of inflation.
Moreover, we assume that the reheating period is short enough and the primary value of $\rho_{\Lambda COBE}$ is given by \cite{chen}
\begin{equation}
\rho_{\Lambda COBE}=1.2\times10^{-9}\Big(\frac{T_{reh}}{10^{16}GeV}\Big)^{4}M_{P}^{4}=4.2T_{reh}^{4}.\label{cobe}
\end{equation}

We remind our assumption that the LECHDE is not coupled to
the inflaton field so that the equation of motion of the inflaton
field is not affected by the existence of the logarithmic entropy-corrected
holographic dark energy. Moreover, we ignore any possible perturbations connected
to the LECHDE model. Therefore, the standard perturbation
equations remain unchanged \cite{vita}.

We know that the perturbation of the longitudinal gauge metric is described as follows \cite{chen}
\begin{equation}
ds^{2}=a^{2}\Big[-\Big(1+2\phi\Big)d\tau^{2}+\Big(1-2\phi\Big)dx^{i}dx^{i}\Big],
\end{equation}
where $\phi$ is the scalar field in the perturbed metric and $\tau$ is
the conformal time. Using  the equation
of motion of the inflaton field (\ref{motion}) and the standard perturbation
equations \cite{vita}, the diagonized equation for $\phi$ in the longitudinal gauge can be obtained as follows \cite{chen}
\begin{equation}
\ddot{\phi}+\Big(H-\frac{2\ddot{\varphi}}{\dot{\varphi}}\Big)\dot{\phi}
+\Big(4\dot{H}-H\frac{2\ddot{\varphi}}{\dot{\varphi}}+\frac{\dot{\varphi}^{2}}{M_P^{2}}\Big)\phi
-\frac{\bigtriangledown^{2}}{a^{2}}\phi=0.\label{phidott}
\end{equation}
 Now, we suppose \cite{chen}
\begin{equation}
u=\frac{\phi}{\dot{\varphi}}.\label{u}
\end{equation}
Then, using Eq. (\ref{phidott}) and (\ref{u}), one can obtain \cite{chen}
\begin{equation}
\frac{d^{2}u_{k}}{d\tau^{2}}+\Big(\frac{-4\epsilon+\delta+\sigma}{\tau^{2}}\Big)u_{k}+k^{2}u_{k}=0,
\end{equation}
where
\begin{equation}
\sigma\equiv\frac{\dot{\varphi}^{2}}{H^{2}M_{P}^{2}}.\label{sig}
\end{equation}
Since in this paper we have considered the presence of the non-perturbative
logarithmic entropy-corrected holographic dark energy model during the inflation,  the comoving curvature
perturbation is no longer conserved. This is due to the fact that the logarithmic entropy-corrected holographic dark energy does not fluctuate while the inflaton field fluctuates, hence the perturbation is not adiabatic. However, we can apply a nearly conserved
quantity \cite{chen}. Using a general differential equation with two small parameters $\epsilon_{1}$ and $\epsilon_{2}$, we  have \cite{tower}
\begin{equation}
\ddot{\phi}+\Big(1+\epsilon_{1}\Big)H\dot{\phi}+\epsilon_{2}H^{2}\phi-\frac{q^{2}}{a^{2}}\phi=0,\label{epsilon}
\end{equation}
where $\epsilon_{1}=-\frac{2\ddot{\phi}}{\dot{\phi}H}$ and $\epsilon_{2}=\frac{4\dot{H}}{H^2}-\frac{2\ddot{\phi}}{\dot{\phi}H}+\frac{\dot{\phi}^2}{M_P^2H^2}$.
One can show that the following quantity is nearly conserved \cite{tower}
\begin{equation}
\mathcal{R}=B\Big[\frac{\dot{\phi}}{H}+\Big(1+\epsilon_{1}
+\frac{\dot{H}}{H^{2}}-\epsilon_{2}\Big)\phi\Big]e^{\int\epsilon_{2}H dt},\label{Rper}
\end{equation}
where $B$ is the constant value. The power spectrum of $\mathcal{R}$ is given
by \cite{tower}
\begin{equation}
\mathcal{P}\equiv\frac{k^{3}}{2\pi^{2}}|\mathcal{R}|^{2},\label{P}
\end{equation}
where $k=aH$. Also, the spectral index is defined as follows \cite{chen}
\begin{equation}
n_{s}-1\equiv\frac{d\ln \mathcal{P}}{d\ln k}|_{k=aH}=-8\epsilon+2\delta+2\sigma.\label{ns}
\end{equation}
%------------------------------------------------
\section{  LECHDE model with  future event horizon cut-off in inflation}
In this section, we investigate the evolution of the logarithmic entropy-corrected holographic dark energy model with the future event horizon cut-off in the
inflation. The future event horizon cut-off is given by
\begin{equation}
R_{h}=a\int_{t}^{\infty}\frac{dt}{a}=a\int_{x}^{\infty}\frac{dx}{aH}.\label{event}
\end{equation}
Taking time derivative of Eq. (\ref{event}) and using Eq. (\ref{event}) one can obtain
\begin{equation}
\dot{R_{h}}=HR_{h}-1.\label{eve}
\end{equation}
Now, using Eq. (\ref{density}) and $L=R_{h}$ we  have
\begin{equation}
\rho_{\Lambda}=3c^{2}M_{P}^{2}R_{h}^{-2}+R_{h}^{-4}\Big[
\alpha \ln(M_{P}^{2}R_{h}^{2})+\beta\Big].\label{densityevent}
\end{equation}
 We can rewrite Eq. (\ref{densityevent}) as follows
\begin{equation}
\rho_{\Lambda}=3c^{2}M_{P}^{2}R_{h}^{-2}\gamma_{\mu},\label{rhoevent}
\end{equation}
where
\begin{equation}
\gamma_{\mu}=1+\frac{1}{3c^2M_{P}^{2}R_{h}^{2}}\Big[
\alpha \ln(M_{P}^{2}R_{h}^{2})+\beta\Big].\label{mu}
\end{equation}
Using Eq. (\ref{rhoevent}) and inserting in Eq. (\ref{Omega}), we have
\begin{equation}
\Omega_{\Lambda}=\frac{c^{2}\gamma_{\mu}}{R_{h}^{2}H^{2}}.\label{fractial}
\end{equation}
In the flat FRW universe, using Eqs. (\ref{varphi}) and (\ref{rhoevent}) for the inflation model and the LECHDE model with  the future event horizon cut-off, the Friedmann equation is given by
\begin{equation}
3M_{P}^{2}H^{2}=\frac{1}{2}\dot{\varphi}^{2}+V(\varphi)+
3c^{2}M_{P}^{2}R_{h}^{-2}\gamma_{\mu}.\label{fridman}
\end{equation}
Taking time derivative of Eq. (\ref{fridman}) and using Eqs. (\ref{motion}),
(\ref{eve}) and (\ref{rhoevent}), leads to
\begin{equation}
-2M_{P}^{2}\dot{H}=\dot{\varphi}^{2}+2c^{2}M_{P}^{2}R_{h}^{-2}\gamma_{\mu}
\Big(1-\frac{1}{R_{h}H}\Big)\Big[2-\frac{1}{\gamma_{\mu}}-\frac{\alpha H^{2}\Omega_{\Lambda}}
{3c^{2}M_{P}^{2}\gamma_{\mu}^{2}}\Big].\label{Hdot}
\end{equation}
Using Eqs. (\ref{fractial}), (\ref{fridman}) and the slow-roll conditions, we can obtain the Friedmann equation as follows
\begin{equation}
\frac{1}{H}=\sqrt{1-\Omega_{\Lambda}}\sqrt{\frac{3M_{P}^{2}}{V}}.\label{HH}
\end{equation}
Taking time derivative Eq. (\ref{Omega}) and using Eqs.  (\ref{Omega}), (\ref{eve}), (\ref{rhoevent}), (\ref{mu}), (\ref{Hdot}), (\ref{HH}) and the slow-roll conditions, we obtain the following differential equation
\begin{equation}
\Omega_{\Lambda}^{'}=-2\Omega_{\Lambda}\Big(1-\Omega_{\Lambda}\Big)
\Big(1-\frac{\sqrt{\Omega_{\Lambda}}}{c\sqrt{\gamma_{\mu}}}\Big)
\Big[2-\frac{1}{\gamma_{\mu}}-\frac{\alpha\Omega_{\Lambda}V}
{9c^{4}M_{P}^{4}\gamma_{\mu}^{2}(1-\Omega_{\Lambda})}\Big],\label{prime}
\end{equation}
where, the prime denotes the derivative with respect to $x=\ln (a)$ and $a$ is the scale factor. Now, by assuming $V=M_{P}^{4}$ \cite{david} and inserting in Eq. (\ref{prime}), we obtain
\begin{equation}
\Omega_{\Lambda}^{'}=-2\Omega_{\Lambda}\Big(1-\Omega_{\Lambda}\Big)
\Big(1-\frac{\sqrt{\Omega_{\Lambda}}}{c\sqrt{\gamma_{\mu}}}\Big)
\Big[2-\frac{1}{\gamma_{\mu}}-\frac{\alpha\Omega_{\Lambda}}
{9c^{4}\gamma_{\mu}^{2}(1-\Omega_{\Lambda})}\Big].\label{pri}
\end{equation}

\begin{figure}[ht]
  \centering
  \includegraphics[width=2in]{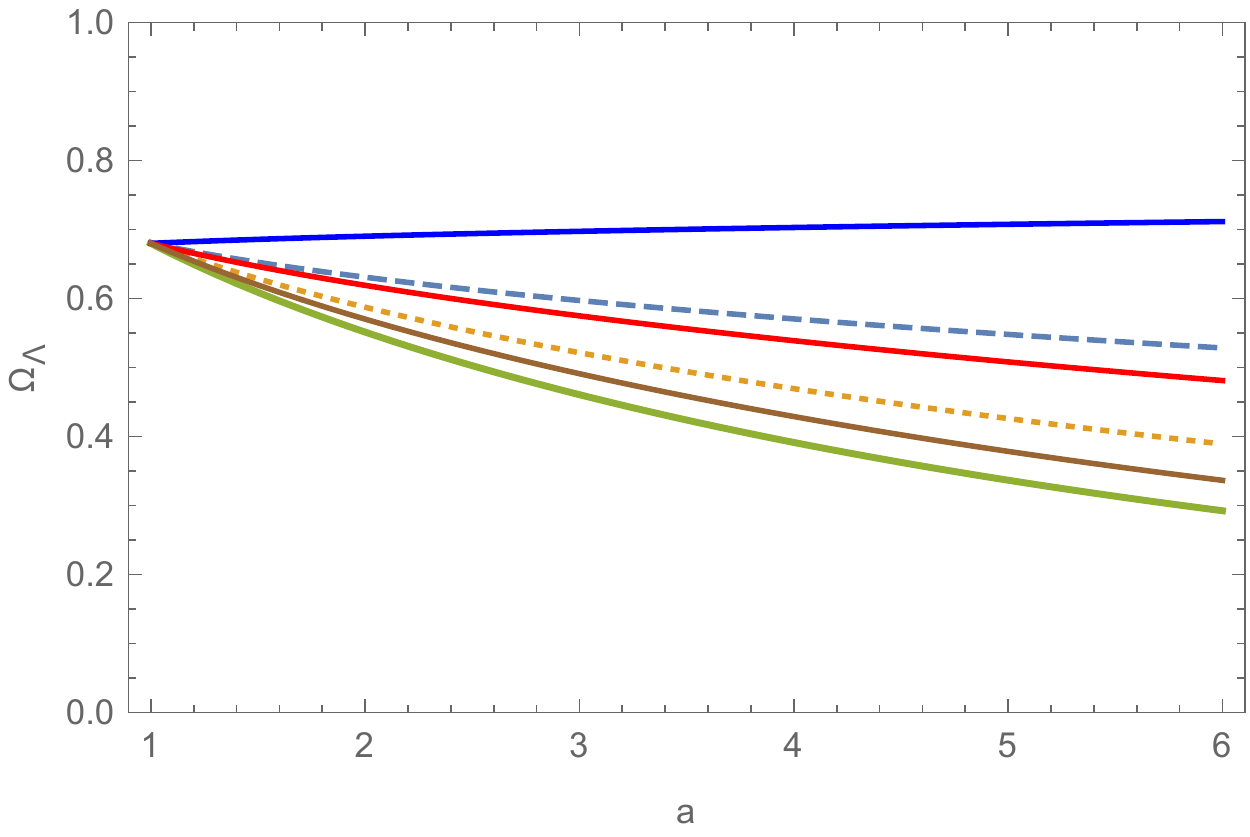}~
    \caption{ Comparison of the evolution of $\Omega_{\Lambda}-a$ between the LECHDE and the HDE models with the future event horizon cut-off.
The dashed, dotted and thick (Green) lines represent the LECHDE model for $c=0.8,1,1.2$, respectively. The Blue, Red and Brown lines indicate the HDE model for $c=0.8,1,1.2$, respectively.}
\label{1}
\end {figure}

This equation has not an analytic solution, however  we have plotted numerically
the evolution of $\Omega_{\Lambda}$ with respect to the scale factor for the logarithmic entropy-corrected
holographic dark energy model with the future event horizon cut-off and the ordinary holographic dark energy model HDE  with $c=0.8,1,1.2$ \cite{chen},
\cite{gong} in the figure (\ref{1}). Note that if $\alpha=\beta=0$ and $\gamma_{\mu}=1$, then Eq. (\ref{pri}) will reduce to Eq. (9) in Ref \cite{chen} and this  means that the LECHDE model will reduce to the HDE model.

\begin{figure}[ht]
\includegraphics[width=2in]{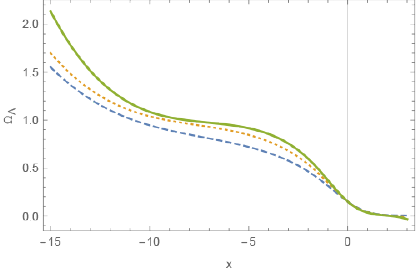}
\caption{The evolution of $\Omega_{\Lambda}$ with respect to $x=\ln (a)$ for the LECHDE model with the future event horizon cut-off. The dashed, dotted and thick lines represent the LECHDE model for $c=0.8,1,1.2$, respectively. Here, neither $a$ nor $\Omega_{\Lambda}$ are not normalized to $a_{0}=1$ and $0<\Omega_{\Lambda}<1$. The normalization is chosen for numerical convenience. For simplicity, the inflaton energy density is assumed to be constant.}
\label{2}
\end{figure}

In this figure, we can see that the evolution of $\Omega_{\Lambda}$ with respect to the scale factor for the LECHDE model with $c=0.8,1,1.2$ is faster than
the evolution of $\Omega_{\Lambda}$ with respect to the scale factor for the HDE model.
In the figure (\ref{2}), we have plotted the evolution of $\Omega_{\Lambda}$ with respect to $x=\ln (a)$ for the logarithmic entropy-corrected
holographic dark energy model for different values of $c=0.8,1,1.2$ \cite{gong}. We
have neglected the change of the inflaton energy density in time for simplicity.
We  see that as  $c$ increases,  the energy density $\Omega_{\Lambda}$ becomes more dominated at earlier times.

Using Eqs. (\ref{cobe}) and (\ref{rhoevent}), we  obtain
\begin{equation}
R_{h}=\frac{5c\times10^{-4}}{M_{P}}\Big(\frac{10^{16}GeV}{T_{reh}}\Big)^{2}\sqrt{\gamma_{\mu}}.\label{Rreh}
\end{equation}
Note that if $\gamma_{\mu}=1$, then Eq. (\ref{Rreh}) will reduce to Eq. (14) in Ref \cite{chen}.

Also, using Eqs. (\ref{delta}), (\ref{epp}), (\ref{phidott}), (\ref{sig}), (\ref{epsilon}), (\ref{Rper}), (\ref{Hdot}) and (\ref{HH}), we  obtain
\begin{equation}
\mathcal{R}=\frac{2M_{P}^{2}H^{2}}{\dot{\varphi}^{2}}\Big(\frac{\dot{\phi}}{H}+\phi\Big)
\exp\Big(2c^{2}\int_{t}^{t_{LS}}\frac{\gamma_{\mu}}{R_{h}^{2}H}\Big[\Big(1-\frac{1}{R_{h}H}\Big)
\Big(2-\frac{1}{\gamma_{\mu}}
-\frac{\alpha\Omega_{\Lambda}}{9c^{4}\gamma_{\mu}^{2}(1-\Omega_{\Lambda})}\Big)\Big]dt\Big),\label{scalar}
\end{equation}
where $t_{LS}$ is the time of the Last Scattering Surface. Using (\ref{P}) and (\ref{scalar}), we also obtain
\begin{equation}
\mathcal{P}=\frac{H^{4}}{4\pi^{2}\dot{\varphi}^{2}}
\exp\Big(4c^{2}\int_{t}^{t_{LS}}\frac{\gamma_{\mu}}{R_{h}^{2}H}\Big[\Big(1-\frac{1}{R_{h}H}\Big)
\Big(2-\frac{1}{\gamma_{\mu}}
-\frac{\alpha\Omega_{\Lambda}}{9c^{4}\gamma_{\mu}^{2}(1-\Omega_{\Lambda})}\Big)\Big]dt\Big).\label{scal}
\end{equation}
Finally, using Eqs. (\ref{delta}), (\ref{epp}), (\ref{sig}), (\ref{ns}), (\ref{Hdot})
and (\ref{HH}), we find
\begin{equation}
n_{s}-1=-4\epsilon+2\delta-\frac{4c^{2}\gamma_{\mu}}{R_{h}^{2}H^{2}}
\Big[\Big(1-\frac{1}{R_{h}H}\Big)\Big(2-\frac{1}{\gamma_{\mu}}
-\frac{\alpha\Omega_{\Lambda}}{9c^{4}\gamma_{\mu}^{2}(1-\Omega_{\Lambda})}\Big)\Big].\label{nnn}
\end{equation}
If $\alpha=\beta=0$ and $\gamma_{\mu}=1$, then Eqs. (\ref{scalar}), (\ref{scal})
and (\ref{nnn}) will reduce to Eqs. (20), (21) and (22), respectively in Ref \cite{chen}.

Now, we  derive the corrections to the spectral index produced by
the logarithmic entropy-corrected holographic dark energy with the future
event horizon cut-off. The slow roll
parameter is given by \cite{chen}
\begin{equation}
\eta\equiv\epsilon+\delta.\label{eta}
\end{equation}
Using Eqs. (\ref{epp}), (\ref{Hdot}) and (\ref{HH}), we have
\begin{equation}
\epsilon=\epsilon_{0}+\frac{c^{2}\gamma_{\mu}}{R_{h}^{2}H^{2}}
\Big[\Big(1-\frac{1}{R_{h}H}\Big)\Big(2-\frac{1}{\gamma_{\mu}}
-\frac{\alpha\Omega_{\Lambda}}{9c^{4}\gamma_{\mu}^{2}(1-\Omega_{\Lambda})}\Big)\Big],\label{eppsilon01}
\end{equation}
where $\epsilon_{0}$ is the main contribution in the inflation models without the logarithmic entropy-corrected holographic dark energy. In the above equation, the correction terms are as follows
\begin{eqnarray}
\frac{c^{2}\gamma_{\mu}}{R_{h}^{2}H^{2}}
\Big[\Big(1-\frac{1}{R_{h}H}\Big)\Big(2-\frac{1}{\gamma_{\mu}}
-\frac{\alpha\Omega_{\Lambda}}{9c^{4}\gamma_{\mu}^{2}(1-\Omega_{\Lambda})}\Big)\Big].\nonumber\\
\end{eqnarray}
Using Eqs. (\ref{nnn}), (\ref{eta}) and (\ref{eppsilon01}), we have
\begin{equation}
n_{s}-1=-6\epsilon_{0}+2\eta-\frac{10c^{2}\gamma_{\mu}}{R_{h}^{2}H^{2}}
\Big[\Big(1-\frac{1}{R_{h}H}\Big)\Big(2-\frac{1}{\gamma_{\mu}}
-\frac{\alpha\Omega_{\Lambda}}{9c^{4}\gamma_{\mu}^{2}(1-\Omega_{\Lambda})}\Big)\Big].\label{terms}
\end{equation}
Here the first and second terms are the standard contributions from the single
field inflaton models. Also, because of the last term in Eq. (\ref{terms}), we can see that the effect of the LECHDE model with the future event horizon cut-off is to make the spectrum redder than that of the HDE model.
Moreover, using the cosmological data \cite{ade} (the correction to $n_s -1$ should be smaller than -0.05), Eq. (\ref{fractial}), and the last term in Eq. (\ref{terms}), we obtain a constraint as follows
\begin{eqnarray}
\frac{2x^{4}}{5y^{2}}
\Big[\Big(1-\frac{x^{2}}{5cy\sqrt{\gamma_{\mu}}}\Big)\Big(2-\frac{1}{\gamma_{\mu}}
-\frac{\alpha x^{4}}{9c^{4}\gamma_{\mu}^{2}(25y^{2}-x^{4})}\Big)\Big]-0.05<0.\label{ccc}
\end{eqnarray}
where $x\equiv\frac{T_{reh}}{10^{16}GeV}$,  $y\equiv\frac{H}{10^{-4}M_{P}}$
and $\gamma_{\mu}$ is given in terms of $x$, $\alpha, \beta$ and $c$ by solving the following equation
\begin{eqnarray}
\gamma_{\mu}=1+\frac{x^4}{75\times 10^8 c^4\gamma_{\mu}}\Big[\alpha \ln \Big(\frac{25\times 10^8 c^2\gamma_{\mu}}{x^4}\Big)+\beta\Big].
\end{eqnarray}
The inequality (\ref{ccc}) constrains the quantities $H, T_{reh}, c$, $\alpha$
and $\beta$ at early universe.
%-----------------------------------------------------------------------------------------------
\section{ LECHDE model with  particle horizon cut-off in inflation}
 The particle horizon cut-off is given by
\begin{equation}
R_{H}=a\int_{0}^{t}\frac{dt}{a}.\label{particle}
\end{equation}
Taking time derivative of Eq. (\ref{particle}) and using Eq. (\ref{particle}) one can obtain
\begin{equation}
\dot{R_{H}}=HR_{H}+1.\label{part}
\end{equation}
Using Eq. (\ref{density}) and $L=R_{H}$ we have
\begin{equation}
\rho_{\Lambda}=3c^{2}M_{P}^{2}R_{H}^{-2}+R_{H}^{-4}\Big[
\alpha \ln(M_{P}^{2}R_{H}^{2})+\beta\Big].\label{densityparticle}
\end{equation}
Also, we can write Eq. (\ref{densityparticle}) as follows
\begin{equation}
\rho_{\Lambda}=3c^{2}M_{P}^{2}R_{H}^{-2}\gamma_{\nu},\label{rhoparti}
\end{equation}
where
\begin{equation}
\gamma_{\nu}=1+\frac{1}{3c^2M_{P}^{2}R_{H}^{2}}\Big[
\alpha \ln(M_{P}^{2}R_{H}^{2})+\beta\Big].\label{nu}
\end{equation}
Using Eq. (\ref{rhoparti}) and inserting in Eq. (\ref{Omega}), we have
\begin{equation}
\Omega_{\Lambda}=\frac{c^{2}\gamma_{\nu}}{R_{H}^{2}H^{2}}.\label{fractial11}
\end{equation}
In the flat FRW universe, using Eqs. (\ref{varphi}) and (\ref{rhoparti}) for the inflation model and the LECHDE model with the particle horizon cut-off, the Friedmann equation is given by
\begin{equation}
3M_{P}^{2}H^{2}=\frac{1}{2}\dot{\varphi}^{2}+V+
3c^{2}M_{P}^{2}R_{H}^{-2}\gamma_{\nu}.\label{fridman1}
\end{equation}
Taking the time derivative of Eq. (\ref{fridman1}) and using Eqs. (\ref{motion}),
(\ref{part}) and (\ref{rhoparti}) yields
\begin{equation}
-2M_{P}^{2}\dot{H}=\dot{\varphi}^{2}+2c^{2}M_{P}^{2}R_{H}^{-2}\gamma_{\nu}
\Big(1+\frac{1}{R_{H}H}\Big)\Big[2-\frac{1}{\gamma_{\nu}}-\frac{\alpha H^{2}\Omega_{\Lambda}}
{3c^{2}M_{P}^{2}\gamma_{\nu}^{2}}\Big].\label{Hdot1}
\end{equation}
Using Eqs. (\ref{fractial11}), (\ref{fridman1}) and the slow-roll conditions, we obtain the Friedmann equation as follows \begin{equation}
\frac{1}{H}=\sqrt{1-\Omega_{\Lambda}}\sqrt{\frac{3M_{P}^{2}}{V}}.\label{HH1}
\end{equation}
Taking  time derivative of Eq. (\ref{Omega}) and using Eqs.  (\ref{Omega}), (\ref{part}), (\ref{rhoparti}), (\ref{nu}), (\ref{Hdot1}), (\ref{HH1}) and the slow-roll conditions, we have
\begin{equation}
\Omega_{\Lambda}^{'}=-2\Omega_{\Lambda}\Big(1-\Omega_{\Lambda}\Big)
\Big(1+\frac{\sqrt{\Omega_{\Lambda}}}{c\sqrt{\gamma_{\nu}}}\Big)
\Big[2-\frac{1}{\gamma_{\nu}}-\frac{\alpha\Omega_{\Lambda}V}
{9c^{4}M_{P}^{4}\gamma_{\nu}^{2}(1-\Omega_{\Lambda})}\Big].\label{prime1}
\end{equation}
 Now, we assume $V=M_{P}^{4}$ \cite{david} and insert in Eq. (\ref{prime1}) to obtain
\begin{equation}
\Omega_{\Lambda}^{'}=-2\Omega_{\Lambda}\Big(1-\Omega_{\Lambda}\Big)
\Big(1+\frac{\sqrt{\Omega_{\Lambda}}}{c\sqrt{\gamma_{\nu}}}\Big)
\Big[2-\frac{1}{\gamma_{\nu}}-\frac{\alpha\Omega_{\Lambda}}
{9c^{4}\gamma_{\nu}^{2}(1-\Omega_{\Lambda})}\Big].\label{pri1}
\end{equation}
Similar to the previous case, in the figure (3),  we have plotted numerically
the evolution of $\Omega_{\Lambda}$ with respect to the scale factor for the logarithmic entropy-corrected
holographic dark energy model with the particle horizon cut-off and the ordinary holographic dark energy model HDE  with $c=0.8,1,1.2$ \cite{chen},
\cite{gong}. Note that if $\alpha=\beta=0$ and $\gamma_{\nu}=1$, then Eq. (50) will reduce to Eq. (34) in Ref \cite{chen} and this  means that the LECHDE model will reduce to the HDE model.
In this figure, we can see that unlike the previous case the evolution of $\Omega_{\Lambda}$ with respect to the scale factor for the LECHDE model with $c=0.8,1,1.2$ is slower than
the evolution of $\Omega_{\Lambda}$ with respect to the scale factor for the HDE model.
\begin{figure}[ht]
  \centering
  \includegraphics[width=2in]{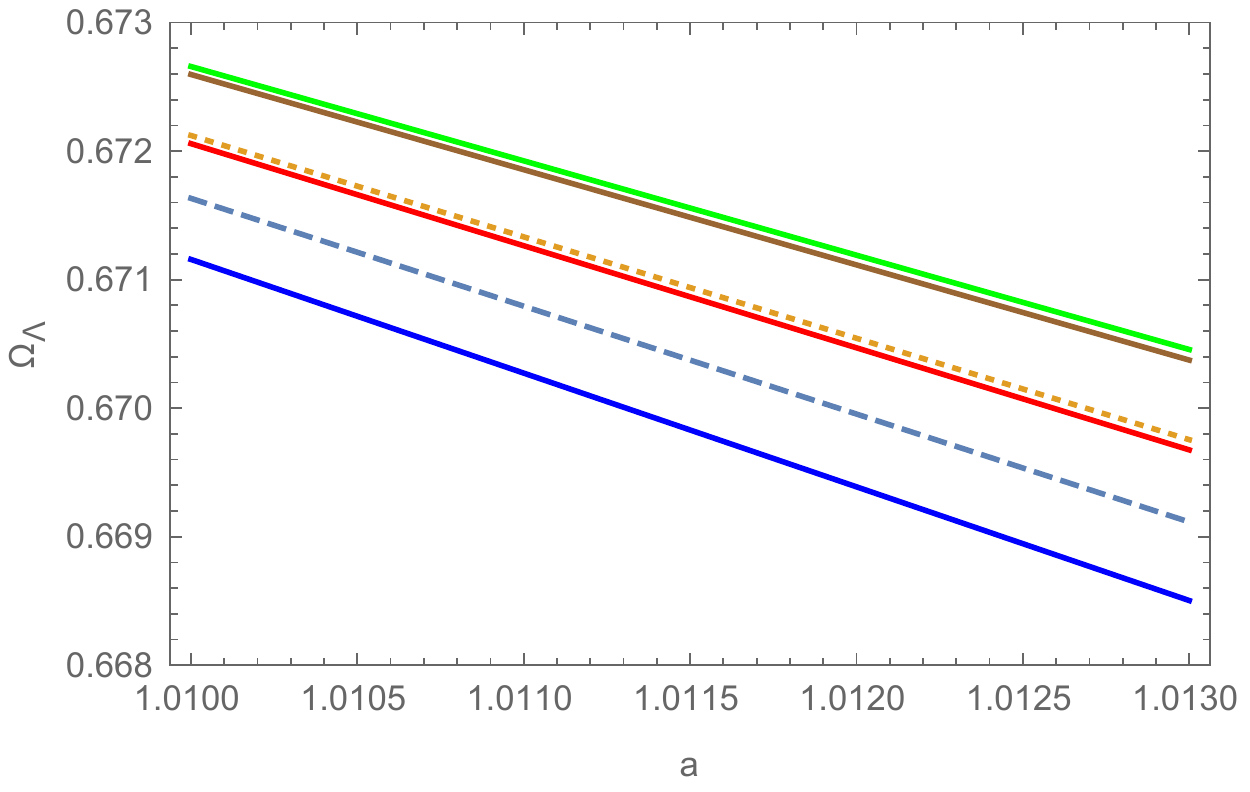}~
    \caption{The comparison evolution $\Omega_{\Lambda}-a$ between the LECHDE
     and the HDE models with the particle horizon cut-off for two models.
The dashed, dotted and thick (Green) lines represent the LECHDE model for $c=0.8,1,1.2$, respectively. The Blue, Red and Brown lines indicate the HDE model for $c=0.8,1,1.2$, respectively.}
\label{4}
\end {figure}
In the figure (4), we have plotted the evolution of $\Omega_{\Lambda}$ with respect to $x=\ln (a)$ for the logarithmic entropy-corrected
holographic dark energy model with $c=0.8,1,1.2$ \cite{gong}. For simplicity,
we have neglected the change of the inflaton energy density with  time.
We  see that as  $c$ increases,  the energy density $\Omega_{\Lambda}$ becomes more dominated at earlier times.
\begin{figure}[ht]
  \centering
  \includegraphics[width=2in]{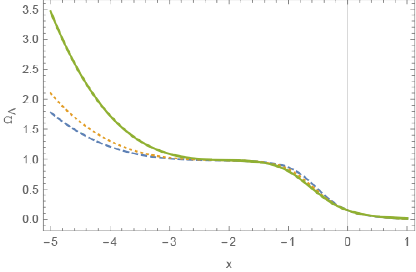}~
    \caption{The evolution $\Omega_{\Lambda}$ respect to $x=\ln (a)$ for the LECHDE model with the particle horizon cut-off. The dashed, dotted and thick lines represent the LECHDE model for $c=0.8,1,1.2$, respectively. Here, neither $a$ nor $\Omega_{\Lambda}$ are not normalized to $a_{0}=1$ and $0<\Omega_{\Lambda}<1$. The normalization is chosen for numerical convenience. For simplicity, the inflaton energy density is assumed to be constant.}
\label{5}
\end {figure}

Using Eqs. (\ref{cobe}) and (\ref{rhoparti}), we obtain
\begin{equation}
R_{H}=\frac{5c\times10^{-4}}{M_{P}}\Big(\frac{10^{16}GeV}{T_{reh}}\Big)^{2}\sqrt{\gamma_{\nu}}.\label{Rreh11}
\end{equation}
Also, using Eqs. (\ref{delta}), (\ref{epp}), (\ref{phidott}), (\ref{sig}), (\ref{epsilon}), (\ref{Rper}), (\ref{Hdot1}) and (\ref{HH1}), we  find
\begin{equation}
\mathcal{R}=\frac{2M_{P}^{2}H^{2}}{\dot{\varphi}^{2}}\Big(\frac{\dot{\phi}}{H}+\phi\Big)
\exp\Big(2c^{2}\int_{t}^{t_{LS}}\frac{\gamma_{\nu}}{R_{H}^{2}H}\Big[\Big(1+\frac{1}{R_{H}H}\Big)
\Big(2-\frac{1}{\gamma_{\nu}}
-\frac{\alpha\Omega_{\Lambda}}{9c^{4}\gamma_{\nu}^{2}(1-\Omega_{\Lambda})}\Big)\Big]dt\Big).\label{scalar1}
\end{equation}
 Using (\ref{P}) and (\ref{scalar1}) yields
\begin{equation}
\mathcal{P}=\frac{H^{4}}{4\pi^{2}\dot{\varphi}^{2}}
\exp\Big(4c^{2}\int_{t}^{t_{LS}}\frac{\gamma_{\nu}}{R_{H}^{2}H}\Big[\Big(1+\frac{1}{R_{H}H}\Big)
\Big(2-\frac{1}{\gamma_{\nu}}
-\frac{\alpha\Omega_{\Lambda}}{9c^{4}\gamma_{\nu}^{2}(1-\Omega_{\Lambda})}\Big)\Big]dt\Big).\label{scal11}
\end{equation}
Using Eqs. (\ref{delta}), (\ref{epp}), (\ref{sig}), (\ref{ns}), (\ref{Hdot1})
and (\ref{HH1}), we have
\begin{equation}
n_{s}-1=-4\epsilon+2\delta-\frac{4c^{2}\gamma_{\nu}}{R_{H}^{2}H^{2}}
\Big[\Big(1+\frac{1}{R_{H}H}\Big)\Big(2-\frac{1}{\gamma_{\nu}}
-\frac{\alpha\Omega_{\Lambda}}{9c^{4}\gamma_{\nu}^{2}(1-\Omega_{\Lambda})}\Big)\Big].\label{nnn1}
\end{equation}
For $\alpha=\beta=0$ and $\gamma_{\nu}=1$, Eqs. (\ref{scalar1}), (\ref{scal11})
and (\ref{nnn1}) will reduce to Eqs. (35), (36) and (37), respectively in Ref \cite{chen}.
Using Eqs. (\ref{epp}), (\ref{Hdot1}) and (\ref{HH1}), we obtain
\begin{equation}
\epsilon=\epsilon_{0}+\frac{c^{2}\gamma_{\nu}}{R_{H}^{2}H^{2}}
\Big[\Big(1+\frac{1}{R_{H}H}\Big)\Big(2-\frac{1}{\gamma_{\nu}}
-\frac{\alpha\Omega_{\Lambda}}{9c^{4}\gamma_{\nu}^{2}(1-\Omega_{\Lambda})}\Big)\Big],\label{eppsilon0101}
\end{equation}
where $\epsilon_{0}$ is the main contribution in the inflation models without the logarithmic entropy-corrected holographic dark energy. In the above equation, the correction terms are as follows
\begin{eqnarray}
\frac{c^{2}\gamma_{\nu}}{R_{H}^{2}H^{2}}
\Big[\Big(1+\frac{1}{R_{H}H}\Big)\Big(2-\frac{1}{\gamma_{\nu}}
-\frac{\alpha\Omega_{\Lambda}}{9c^{4}\gamma_{\nu}^{2}(1-\Omega_{\Lambda})}\Big)\Big].\nonumber\\
\end{eqnarray}
Using Eqs. (\ref{eta}), (\ref{nnn1}) and (\ref{eppsilon0101}), we find
\begin{equation}
n_{s}-1=-6\epsilon_{0}+2\eta-\frac{10c^{2}\gamma_{\nu}}{R_{H}^{2}H^{2}}
\Big[\Big(1+\frac{1}{R_{H}H}\Big)\Big(2-\frac{1}{\gamma_{\nu}}
-\frac{\alpha\Omega_{\Lambda}}{9c^{4}\gamma_{\nu}^{2}(1-\Omega_{\Lambda})}\Big)\Big].\label{terms11}
\end{equation}
As in the previous case,  we can see that the effect of the LECHDE model with the particle horizon cut-off is to make the spectrum redder.
Using the cosmological data \cite{ade} (the correction to $n_s -1$ should be smaller than -0.05), Eq. (\ref{fractial11}), and the last term in Eq. (\ref{terms11}), we obtain a constraint as follows
\begin{eqnarray}
\frac{2x^{4}}{5y^{2}}
\Big[\Big(1+\frac{x^{2}}{5cy\sqrt{\gamma_{\nu}}}\Big)\Big(2-\frac{1}{\gamma_{\nu}}
-\frac{\alpha x^{4}}{9c^{4}\gamma_{\nu}^{2}(25y^{2}-x^{4})}\Big)\Big]-0.05<0.\label{cccc}
\end{eqnarray}
where $x\equiv\frac{T_{reh}}{10^{16}GeV}$,  $y\equiv\frac{H}{10^{-4}M_{P}}$
and $\gamma_{\nu}$ is given in terms of $x$, $\alpha, \beta$ and $c$ by solving the following equation
\begin{eqnarray}
\gamma_{\nu}=1+\frac{x^4}{75\times 10^8 c^4\gamma_{\nu}}\Big[\alpha \ln \Big(\frac{25\times 10^8 c^2\gamma_{\nu}}{x^4}\Big)+\beta\Big].
\end{eqnarray}
The inequality (\ref{cccc}) constrains the quantities $H, T_{reh}, c$, $\alpha$
and $\beta$ at early universe.

%-----------------------------------------------------------------------------------------------
\section{ LECHDE model with  Hubble cut-off in inflation}
 The hubble cut-off is given by
\begin{equation}
L=H^{-1}.\label{hubble}
\end{equation}

Using Eqs. (\ref{density}) and (\ref{hubble}) we find
\begin{equation}
\rho_{\Lambda}=3c^{2}M_{P}^{2}H^{2}+H^{4}\Big[
\alpha \ln(M_{P}^{2}H^{-2})+\beta\Big].\label{densityhubble}
\end{equation}
Also, we can write Eq. (\ref{densityhubble}) as follows
\begin{equation}
\rho_{\Lambda}=3c^{2}M_{P}^{2}H^{2}\gamma_{\theta},\label{rhohubble}
\end{equation}
Where
\begin{equation}
\gamma_{\theta}=1+\frac{H^{2}}{3c^2M_{P}^{2}}\Big[
\alpha \ln(M_{P}^{2}H^{-2})+\beta\Big].\label{theta}
\end{equation}
Using Eq. (\ref{rhohubble}) and inserting in Eq. (\ref{Omega}), we have
\begin{equation}
\Omega_{\Lambda}=c^{2}\gamma_{\theta}.\label{fractialhubble}
\end{equation}
In the flat FRW universe, using Eqs. (\ref{varphi}) and (\ref{rhohubble}) for the inflation model and the LECHDE model with the Hubble cut-off, the Friedmann equation is given by
\begin{equation}
3M_{P}^{2}H^{2}=\frac{1}{2}\dot{\varphi}^{2}+V+
3c^{2}M_{P}^{2}H^{2}\gamma_{\theta}.\label{fridmanhubble}
\end{equation}
Now, taking time derivative of Eq. (\ref{theta}) and using Eq. (\ref{theta})
yields
\begin{equation}
\dot{\gamma_{\theta}}=\dot{H}\Big[\frac{2(\gamma_{\theta}-1)}{H}-\frac{2\alpha
H}{3c^{2}M_{P}^{2}}\Big].\label{gamateta}
\end{equation}
Also, taking time derivative of Eq. (\ref{fridmanhubble}) and using Eqs. (\ref{motion}),
(\ref{rhohubble}) and (\ref{gamateta}) leads to
\begin{equation}
-2M_{P}^{2}\dot{H}\Big[1-c^{2}\Big(2\gamma_{\theta}-1\Big)+\frac{\alpha H^{2}}{3M_{P}^{2}}\Big]
=\dot{\varphi}^{2}.\label{Hdothubbel}
\end{equation}
Using Eqs. (\ref{fractialhubble}), (\ref{fridmanhubble}) and the slow-roll conditions, the Friedmann equation is obtained as follows
\begin{equation}
\frac{1}{H}=\sqrt{1-\Omega_{\Lambda}}\sqrt{\frac{3M_{P}^{2}}{V}}.\label{HH1hubble}
\end{equation}
Using Eqs. (\ref{delta}), (\ref{epp}), (\ref{phidott}), (\ref{sig}), (\ref{epsilon}), (\ref{Rper}), (\ref{Hdothubbel}) and (\ref{HH1hubble}), we obtain
\begin{equation}
\mathcal{R}=\frac{2M_{P}^{2}H^{2}}{\dot{\varphi}^{2}}
\Big(\frac{\dot{\phi}}{H}+\phi\Big)\exp\Big[2\int_{t}^{t_{LS}}\frac{dH}{H}
\Big(c^{2}(2\gamma_{\theta}-1)-
\frac{\alpha}{9(1-\Omega_{\Lambda})}\Big)\Big].\label{scalar1hubble}
\end{equation}
 Using (\ref{P}) and (\ref{scalar1hubble}), we find
\begin{equation}
\mathcal{P}=\frac{H^{4}}{4\pi^{2}\dot{\varphi}^{2}}
\exp\Big[4\int_{t}^{t_{LS}}\frac{dH}{H}\Big(c^{2}(2\gamma_{\theta}-1)-
\frac{\alpha}{9(1-\Omega_{\Lambda})}\Big)\Big].\label{scal11hubble}
\end{equation}
And, using Eqs. (\ref{delta}), (\ref{epp}), (\ref{sig}), (\ref{ns}), (\ref{Hdothubbel})
and (\ref{HH1hubble}), we have
\begin{equation}
n_{s}-1=2\delta-\epsilon\Big[4+4c^{2}\Big(2\gamma_{\theta}-1\Big)-
\frac{\alpha}{9(1-\Omega_{\Lambda})}\Big].\label{nnn1hubble}
\end{equation}
We assume $V=M_{P}^{4}$ \cite{david} and insert Eq. (\ref{HH1hubble}) in Eq. (\ref{theta}) to obtain
\begin{equation}
\gamma_{\theta}=1+\frac{1}{9c^{2}\Big(1-\Omega_{\Lambda}\Big)}\Big[\ln\Big(3-3\Omega_{\Lambda}\Big)+1\Big].\label{gamatee}
\end{equation}
In the $\Omega_{\Lambda}\longrightarrow0$ limit, Eq. (\ref{gamatee})
yields \begin{equation}
\gamma_{\theta}=1+\frac{0.233}{c^{2}}.\label{gamatee1}
\end{equation}
Now in the $\Omega_{\Lambda}\longrightarrow0$ limit and using Eq. (\ref{gamatee1}),
Eqs. (\ref{scalar1hubble}), (\ref{scal11hubble}) and (\ref{nnn1hubble}) will
change as follows
\begin{equation}
\mathcal{R}=\frac{2M_{P}^{2}H^{2+2c^{2}+0.71}}{M^{-2c^{2}-0.71}\dot{\varphi}^{2}}
\Big(\frac{\dot{\phi}}{H}+\phi\Big),\label{hubblemath}
\end{equation}
\begin{equation}
\mathcal{P}=\frac{H^{4+4c^{2}+1.42}}{4\pi^{2}M^{4c^{2}+1.42}\dot{\varphi}^{2}},\label{hubblep}
\end{equation}
\begin{equation}
n_{s}-1=-\epsilon\Big(4+4c^{2}+1.753\Big)+2\delta,\label{hubblens}
\end{equation}
where $M$ is a constant with the dimension of energy.
We compare  Eqs. (\ref{hubblemath}), (\ref{hubblep}) and (\ref{hubblens})
with Eqs. (42), (43) and (44) in Ref \cite{chen}. Then, we can write the
above equations as follows
\begin{equation}
\mathcal{R}_{LECHDE}=\mathcal{R}_{HDE}\frac{H^{0.71}}{M^{-0.71}},\label{hubblemath1}
\end{equation}
\begin{equation}
\mathcal{P}_{LECHDE}=\mathcal{P}_{HDE}\frac{H^{1.42}}{M^{1.42}},\label{hubblep1}
\end{equation}
\begin{equation}
(n_{s}-1)_{LECHDE}={(n_{s}-1)}_{HDE}-1.753\epsilon~. \label{hubblens11}
\end{equation}
It is seen that in the $\Omega_{\Lambda}\longrightarrow0$ limit the effect of the LECHDE model with the Hubble cut-off is to make the spectrum redder than that of the HDE model.
%-----------------------------------------------------------------------------------------------
\section{Concluding remarks}\label{Con}
{In this work, we have investigated the inflation by logarithmic entropy-corrected holographic dark energy LECHDE model for different cut-offs. We have assumed that the inflaton field does not couple to the logarithmic entropy-corrected holographic dark energy and hence it  is not affected by the existence of the logarithmic entropy-corrected holographic dark energy. Also, we have assumed that the LECHDE model depends on the background and it does not create the perturbations. Therefore, the standard perturbation
equations remain  unchanged. We have also assumed that the reheating period occurs immediately after the inflation
period. Considering these assumptions,  we have compared our results for the LECHDE model with the results of the HDE
model obtained in \cite{chen}.}
{We have found that for  the future event horizon cut-off (see figure (\ref{1})),  the evolution of $\Omega_{\Lambda}$ with respect to the scale factor for the LECHDE model is faster than that of the HDE
model. Also, in the evolution of $\Omega_{\Lambda}$ with respect to $x=\ln (a)$, we obtained that as  $c$ increases, the energy density $\Omega_{\Lambda}$ becomes more dominated at earlier times  for the LECHDE model 
compared with the HDE model.}
{For the particle horizon cut-off (see figure (\ref{4})), we have found that the evolution of $\Omega_{\Lambda}$ with respect to the scale factor for the LECHDE model is slower than that of the HDE model. Also, in the evolution of $\Omega_{\Lambda}$ with respect to $x=\ln (a)$, we obtained that as  $c$ increases, the energy density $\Omega_{\Lambda}$ becomes more dominated at earlier times  for the LECHDE model compared with the HDE model.}

We have derived the corrections to the spectral index produced by the LECHDE model with the event future horizon, the particle horizon and the hubble horizon cut-offs, and found that the effect of the LECHDE model for all three
cut offs is making the spectrum redder than the HDE model.
The requirement of consistency with the observational data in LECHDE model of inflation, constrains the reheating temperature
and Hubble parameter by one parameter of holographic dark energy and two
new parameters of logarithmic corrections, compared to the HDE model.

%-----------------------------------------------------------------------------------------------
%\section*{Acknowledgment}
%This work has been supported by a grant/research fund number
%$217/D/10976$ from Azarbaijan Shahid Madani University.
%-----------------------------------------------------------------------------------------------
%       REFERENCE LIST
%---------------------------------------------------------------------------------------


\begin{thebibliography}{9}
\bibitem{Riess} A. G. Riess, et al., Astron. J. 116 (1998) 1009;
S. Perlmutter, et al., Astrophys. J. 517 (1999) 565;
P. de Bernardis, et al., Nature 404 (2000) 955;
S. perlmutter, et al., Astrophys. J. 598 (2003) 102.
\bibitem{copeland} E. J. Copeland, M. Sami, S. Tsujikawa, IJMPD, 15 (2066)
1753.
\bibitem{Swein} S. Weinberg, Reviews of modern physics, 61 (1989) 1.
\bibitem{pad} T. Padmanabhan, Phys. Rept. 380 (2006) 235.
\bibitem{Li1}A. G. Cohen, D. B. Kaplan, and A. E. Nelson, Phys. Rev. Lett. 82, (1999) 4971; S. D. H. Hsu, Phys. Lett. B 594, (2004) 13. 
\bibitem{Li2}M. Li, Phys. Lett. B 603, (2004) 1.
\bibitem{wei} H. Wei, R. G. Cai, Phys. Lett. B 660 (2008) 113.
\bibitem{setare} Y. F. Cai, E. N. Saridakis, M. R. Setare, J. Q. Xia, Phys.
Rept. 493 (2010) 1.
\bibitem{msetare} M. R. Setare, Phys. Lett. B 653 (2007) 116.
\bibitem{MRsetare} M. R. Setare, J. sadeghi, A. R. Amani, Phys. Lett. B 673
(2009) 241.
\bibitem{suss} L. Susskind, J. Math. Phys. 36 (1995) 6377;
S. Nojiri, S. D. Odintsov, Gen. Rel. Grav. 38 (1285) 1285;
K. Bamba, S. Capozziello, S. D. Odintsov, Astrophys. Space Sci. 342 (2012)
155.
\bibitem{wald} R. M. Wald, Phys. Rev. D 48 (1993) 3427.
\bibitem{Maj} R. Banerjee, B. R. Majhi, Phys. Lett. B 662 (2008) 62;
R. Banerjee, B. R. Majhi, JHEP 06 (2008) 095;
J. Zhang, Phys. Lett. B 668 (2008) 353.
\bibitem{cai} Y. F. Cai, J. Liu, H. Li, Phys. Lett. B 690 (2010) 213.
\bibitem{WWi} H. Wei, Commun. Theor. Physics, 52 (2009) 743.
\bibitem{chen} B. Chen, M. Li, Y. Wang, Nucl. Phys. B 774 (2007) 256.
\bibitem{david} D. H. Lyth, A. Riotto, Phys. Rept. 314 (1999) 1.
\bibitem{gong} Q. G. Huang, Y. Gong, JCAP 0408 (2004) 006;
H. C. Kao, W. L. Lee, F. L. Lin, Phys. Rev. D 71 (2005) 123518;
X. Zhang, Int. J. Mod. Phys. D 14 (2005) 1597;
X. Zhang, F. Q. Wu, Phys. Rev. D 72 (2005) 043524;
Z. Chang, F. Q. Wu, X. Zhang, Phys. Lett. B 633 (2006) 14;
X. Zhang, Int. J. Mod. Phys. D 74 (2006) 103505.
\bibitem{vita} V. F. Mukhanov, H. A. Feldman, R. H. Brandenberger, Phys.
Rept. 215 (1992) 203.
\bibitem{tower} B. Chen, M. Li, T. Wang, Y. Wang, Mod. Phys. Lett. A 22 (2007)
1987.
\bibitem{ade}  P. A. R. Ade, et al., Astron. Astrophys. 571 (2014) A 22.



\end{thebibliography}
\end{document}